\documentclass[a4paper,11pt]{article}
\usepackage{jheppub} 
\usepackage{adjustbox}
\usepackage{slashed}
\usepackage{tikz}

\title{Exploring two dimensional $\mathbb{Z}_2$ invariant phases with time reversal symmetry and their transitions with topological operations}

\author{Ryan C. Spieler}
\affiliation{University of Texas, Austin, Physics Department, Austin, TX, 78712, USA}

\emailAdd{rcspieler@utexas.edu}

\abstract{We use various topological operations to systematically study phase transitions between theories with $\mathbb{Z}_2$ and time reversal symmetry in two spacetime dimensions.  The phases (and accompanying CFTs) we consider come in two types - bosonic phases that are defined on unorientable manifolds and fermionic phases that are sensitive to a $\text{Pin}^-$ structure.  In both cases, our analysis leads to eight phase diagrams, with the two sets of eight connected by fermionization/bosonization.  Starting from a seed CFT, we obtain the CFT that governs each transition.  Many of these exhibit symmetry enriched criticality.  In addition to showing many symmetry enriched CFTs in their natural habitats, our work discusses the fermionic analogs of the $\mathbb{Z}_2$ bosonic operations, which we have not seen discussed in the literature.}

\begin{document}
\maketitle
\flushbottom
\section{Introduction}
Symmetry is probably the dominant organizing force in theoretical physics.  In recent years, its role has only expanded with its definition, which now includes all topological defects in a quantum field theory, see i.e. \cite{Gaiotto:2014kfa,Cordova:2022ruw,McGreevy:2022oyu,Brennan:2023mmt,Schafer-Nameki:2023jdn,Shao:2023gho,Bhardwaj:2023kri}.  Among its many uses, symmetry characterizes phases of matter and the transitions between them.  This is an old idea, dating back at least to Landau, but has recently been pursued in a new light.  We begin by describing the phase in question by its low energy quantum field theory.  The key to the strategy is to couple the theory in question to a background for the symmetry.  In this paper, the symmetries in question are finite, invertible, zero-form symmetries.  This enables one to track the action of various topological manipulations, such as gauging the symmetry and stacking with an invertible topological theory \footnote{An invertible field theory is one with a single state.  These are the low energy descriptions of short range entangled phases of matter.}.  These generally result in a new field theory that describes the low energy limit of a distinct phase.  These operations organize field theories in an \textit{orbifold groupoid} \cite{Gaiotto:2020iye} consisting of field theories connected by invertible topological operations \footnote{For the reader not \textit{yet} comfortable with abstract nonsense, a groupoid is a special kind of category.  A category consists of objects connected by maps called morphisms.  A groupoid is a category whose morphisms are all isomorphisms - that is to say all the maps between objects are invertible.  This is precisely the sort of structure formed by the collection of theories connected by topological manipulations.  We will address our preference of this term to the commonly used \textit{duality web} later.}.  Systematically studying the orbifold groupoid for a seed theory allows one to identify all gapped phases with a given symmetry.  Additionally, by beginning with a CFT that describes a transition between two gapped phases, one can obtain many CFTs connected by topological manipulations \cite{Karch:2019lnn,Karch:2025jem}.  Moreover, one can use this to uncover dualities \footnote{We use the term duality to mean an order two map from a theory to itself.  It is often used more loosely to mean any order two map between theories.  For instance, gauging a $\mathbb{Z}_2$ symmetry is often referred to as a duality.} that allow one to write the same CFT in multiple guises \footnote{Our treatment here is ahistorical.  \cite{Karch:2019lnn} was inspired by work on webs of 3d dualities discussed in i.e. \cite{Karch:2016sxi,Seiberg:2016gmd,Senthil:2018cru,Turner:2019wnh}. 
 These are also based on topological operations that are discussed in \cite{Witten:2003ya}.}!  This procedure has an encoding in a topological field theory of one higher dimension, in which the various gapped phases are specified by pairs of topological boundaries - one of which tracks the symmetry- and the maps between them are changes in the boundary conditon that does not track the symmetry.  Gapless theories are more general boundary conditions.  This perspective has lead to an explosion of work (see i.e. \cite{TachikawaNotes,Gaiotto:2020iye,Kaidi:2022cpf,Kaidi:2023maf,Zhang:2023wlu,Antinucci:2023ezl,Cordova:2023bja,Moradi:2022lqp,Bhardwaj:2023ayw,Bhardwaj:2023idu,Bhardwaj:2023fca,Bhardwaj:2023bbf,Bhardwaj:2024qrf,Bhardwaj:2024ydc,Huang:2023pyk,Kong:2020cie,Chatterjee:2022kxb,Chatterjee:2022tyg}) and is very useful for non-invertible symmetries for which the bulk may well be more familiar than the boundary.  We will not utilize this perspective here, and will instead work dierctly in two dimensions.

 In this paper, we study two dimensional theories that possess both a global $\mathbb{Z}_2$ symmetry and time-reversal symmetry.  Studies of theories of this sort appear in i.e. \cite{Kapustin:2014tfa,Kapustin:2014dxa,Gaiotto:2015zta,Thorngren:2018bhj,Bhardwaj:2016dtk,Debray:2018wfz,Turzillo:2018ynq,Kobayashi:2019xxg,Kaidi:2019tyf,Bhardwaj:2020ymp,Barkeshli:2023bta,Turzillo:2023yyr,Rey:2025jno}.  We will never gauge time-reversal symmetry, since doing so would involve doing lower dimensional quantum gravity.  We will of course gauge the global $\mathbb{Z}_2$ symmetry and allow stacking with various invertible phases.  The theories we study come in two types.  The first, which we call bosonic, do not depend on any sort of tangential structure.  We can gauge the $\mathbb{Z}_2$ symmetry, or stack with either of two SPTs.  The second, which we call fermionic, depend on a $\text{Pin}^-$ structure, which is an analog of Spin structure that can be defined on any closed two-manifold regardless of whether it is oriented.  The operations involve stacking the Haldane phase, shifting the $\text{Pin}^-$ structure by the background for time-reversal symmetry, and shifting the $\text{Pin}^-$ structure by the background for time-reversal symmetry than stacking the invertible field theory defined by the $\text{Pin}^-$ structure.  

 The remainder of this paper is structured as follows.  In section two, we study bosonic theories.  We begin by introducing Stiefel-Whitney classes, since they are related to backgrounds for time-reversal symmetry.  We then  catalog all possible gapped phases with $\mathbb{Z}_2 \times \mathbb{Z}_2^T$ symmetry and studying the maps that combine with them to form the orbifold groupoid.  We then apply this to phase diagrams involving two phases and obtain all phase diagrams that can be obtained from the trivial-SSB phase diagrams by the operations we discuss.  We conclude the discussion of bosonic phases by obtaining the web of CFTs that govern the aformentioned phase diagrams by starting with the Ising CFT and applying the topological operations to mimic the generation of the phase diagrams.  Many of these CFTs exhibit symmetry enriched criticality / are gapless SPT phases \cite{Verresen:2019igf,Thorngren:2020wet,Li:2022jbf,Wen:2022tkg,Li:2023ani,Wen:2023otf}.  In section three, we study fermionic theories.  We begin by introducing the Arf-Brown-Kervaire (ABK) invariant that is associated to a $\text{Pin}^-$ structure and discuss its use in fermionizing bosonic theories.  We then enumerate all gapped phases with $\mathbb{Z}_2^F \times \mathbb{Z}_2^T$ symmetries and the operations that connect them.  We then repeat what we do for bosonic phases, obtaining webs of fermionic phase diagrams and accompanying CFTs, many of which exhibit symmetry enriched criticality.  In addition to showing many symmetry enriched CFTs in their natural habitats, our work discusses the fermionic analogs of the $\mathbb{Z}_2$ bosonic operations, which we have not seen discussed in the literature.

\section{Bosonic Web}
In this section, we discuss phase transitions between different gapped phases with $\mathbb{Z}_2 \times \mathbb{Z}_2^T$ symmetry.  We begin by introducing the first two Stiefel-Whitney classes of the tangent bundle over the spacetime manifold.  The first serves as a background for time-reversal symmetry.  We then discuss the different gapped phases with this symmetry and the maps between them.  Following \cite{Gaiotto:2020iye}, we refer to such a structure as an orbifold groupoid.  Then, starting with the trivial to SSB phase diagram, we apply the maps in the orbifold groupoid to exhaust all phase diagrams obtained from this by topological manipulations.  We then repeat this process for the CFTs that control the phase diagrams.
\subsection{Time Reversal Symmetry and Stiefel-Whitney}
To examine the effect of symmetries and their manipulations on a quantum field theory, we couple to backgrounds.  $A \in H^1(X;\mathbb{Z}_2)$ is the background for a $\mathbb{Z}_2$ symmetry.  The first Stiefel-Whitney class of $TX$ is the appropriate background for time-reversal symmetry \cite{Kapustin:2014tfa} \footnote{Time reversal backgrounds are also discussed in \cite{Chen:2014xhe} from the point of view of tensor network states.}.  We denote it by $w_1 \in H^1(X;\mathbb{Z}_2)$.  In this section, we explain why this is an appropriate background for time-reversal symmetry.  It is well known the $w_1$ obstructs one's ability to orient a manifold.  In detail, this means that $PD(w_1)$ is a locus of lines across which the orientation of the manifold flips.  This is precisely how a background for a symmetry works!  

A non-vanishing $w_1$ has a few interesting consequences for calculations on a triangulated manifold.  The most dramatic is that $A \cup A$ does not vanish.  Instead, it satisfies:
\begin{equation}
    A \cup A = A \cup w_1.
\end{equation}
The second Stiefel-Whitney class $w_2 \in H^2(X;\mathbb{Z}_2)$ famously obstructs one assigning a spin structure to a manifold.  On closed 2-manifolds, it is related to the first Stiefel-Whitney class by the formula 
\begin{equation}
    w_1 \cup w_1 = w_2,
\end{equation}
which implies that all orientable 2-manifolds admit a spin structure.
\subsection{Gapped Phases and Topological Operations}
We now discuss the gapped phases compatible with a $\mathbb{Z}_2$ symmetry and time reversal.  The first two are the usual suspects, the trivial and SSB phases, whose parition functions are
\begin{equation}
    Z_{\text{Tri}}[A] = 1,
\end{equation}
and
\begin{equation}
    Z_{\text{SSB}}[A] = \delta(A),
\end{equation}
respectively \footnote{To understand this partition function note that the background for a symmetry is a mesh of symmetry defects.  As one crosses a defect labeled by $g$, one is acted upon by $g$.  This would map between different vacua in the SSB phase, which is forbidden.   The role of the delta function is to turn off such backgrounds locking us in a vacuum.  Note that the formula we write is intended for use on any 2-manifold.  In finite volume, there can be tunneling between distinct vacua.  We work at energy scales much lower than the scales of the tunneling.  Note that even though we have followed the usual convention of writing this as a Dirac delta, it is really a Kronecker delta, see footnote 2 of \cite{Karch:2025jem}.}.  There is a variant of the SSB phase involving the first Stiefel-Whitney class \footnote{Note that this partition function forces $\mathbb{Z}_2$ backgrounds to be time-reversal backgrounds rather than destroying them outright.}:
\begin{equation}
    Z_{\text{SSB}'}[A,w_1] = \delta(A+w_1),
\end{equation}
as well as an SPT phase involving $A$ and $w_1$:
\begin{equation}
    Z_{\text{SPT}}[A,w_1] = (-1)^{\int A \cup w_1},
\end{equation}
as well as the Haldane phase, which is an SPT for time-reversal symmetry
\begin{equation}
    Z_{\text{Hald}}[w_1] = (-1)^{\int w_1 \cup w_1} = (-1)^{\int w_2}.
\end{equation}
The Haldane phase can be stacked with many things, yielding the partition functions:
\begin{equation}
    Z_{\text{Hald+SSB}}[A,w_1] = \delta(A) (-1)^{\int w_2},
\end{equation}
\begin{equation}
    Z_{\text{Hald+SPT}}[A,w_1] = (-1)^{\int (w_2 + A\cup w_1)},
\end{equation}
and
\begin{equation}
    Z_{\text{Hald+SSB'}}[A,w_1] = \delta(A+w_1) (-1)^{\int w_2}.
\end{equation}
Note that $Z_{\text{SSB'+SPT}} = Z_{\text{Hald}}$ since $w_1 \cup w_1 = w_2$ on a 2-manifold.  To map between phases, we employ the following topological manipulations:
\begin{equation}
    O: Z[A,w_1,w_2] \mapsto Z[\Check{A},w_1,w_2] = \sum_{a \in H^1(X;\mathbb{Z}_2)} (-1)^{\int a \cup \Check{A}} Z[a,w_1,w_2],
\end{equation}
\begin{equation}
    S_1: Z[A,w_1,w_2] \mapsto Z[A,w_1,w_2] (-1)^{\int A \cup w_1},
\end{equation}
\begin{equation}
    S_2: Z[A,w_1,w_2] \mapsto Z[A,w_1,w_2] (-1)^{\int w_2}.
\end{equation}
We see that the first gauges the $\mathbb{Z}_2$ symmetry \footnote{Throughout this paper, any sums over elements of $H^1(X;\mathbb{Z}_2)$ are implicitly normalized by $\vert H^1(X;\mathbb{Z}_2)\vert^{-1/2}$.  The convenience of this normalization is discussed in \cite{Karch:2019lnn}.}, the second stacks the SPT, and the third stacks the Haldane chain.  Working out the action of these maps on the gapped phases is straightforward.  The results are in table \ref{bosorbgrp}.
\begin{table}[]
    \centering
    \begin{adjustbox}{width=\linewidth}
    \begin{tabular}{|c|c|c|c|c|c|c|c|c|}
    \hline
    $\mathcal{T}$ & Tri & SSB & SSB' & SPT & Hald & Hald+SSB & Hald+SPT & Hald+SSB'  \\
    \hline
    $O(\mathcal{T})$ & SSB & Tri & SPT  & SSB' & Hald+SSB & Hald & Hald+SSB' & Hald+SPT \\
    \hline
    $S_1(\mathcal{T})$ & SPT & SSB & Hald + SSB' & Triv & Hald+SPT  & Hald+SSB & Hald & SSB' \\ 
    \hline
    $S_2(\mathcal{T})$ & Hald & Hald+SSB & Hald+SSB' & Hald+SPT & Triv & SSB & SPT & SSB' \\
    \hline
    \end{tabular}
    \end{adjustbox}
    \caption{The action of the topological manipulations $O$ (gauging), $S_1$ (stacking the first SPT), and $S_2$ (stacking the Haldene chain) on a gapped theory $\mathcal{T}$.  The various choices of $\mathcal{T}$ are as described in the text.}
    \label{bosorbgrp}
\end{table}

Let us briefly discuss an analog of the Kennedy-Tasaki transformation.  For theories with $\mathbb{Z}_2\times \mathbb{Z}_2$ symmetry, this maps from the phase that breaks the entire $\mathbb{Z}_2\times \mathbb{Z}_2$ symmetry to the nontrivial $\mathbb{Z}_2\times \mathbb{Z}_2$ SPT.  We can do something analogous:
\begin{equation}
    Z_{\text{SPT}}[A,w_1] = S_1 \circ O \circ S_1 (Z_{\text{SSB}}[A]).
\end{equation}
Note that exchanging the roles of $S_1$ and $O$ does not yield the SPT, in contrast to the fact that we can exchange gauging and stacking at will when performing the Kennedy-Tasaki transformation.  This is related to the fact that gauging the entire $\mathbb{Z}_2\times\mathbb{Z}_2$ symmetry maps the SPT to itself.  Here, we only gauge the $\mathbb{Z}_2$ global symmetry, since a discussion of gauging time-reversal symmetry would involve doing quantum gravity \footnote{A recent discussion of this appears in \cite{McNamara:2022lrw,Harlow:2023hjb}}.  If do so, maybe a more fully analogous transformation involving gauging both $\mathbb{Z}_2$ and time-reversal symmetry can be fleshed out.
\subsection{Phase Diagrams}
We will denote a phase diagram by $\mathcal{PD}$.  In order to get a web of phase diagrams, we need to start with a seed.  Our seed will be
\begin{equation}
    \mathcal{PD}_1 = (\text{Tri},\text{SSB})= O(\mathcal{PD}_1) = S_1(\mathcal{PD}_2) = S_2(\mathcal{PD}_3).
\end{equation}
We will not distinguish between the above pair and the pair with the opposite order.  Thus, $O$ maps $\mathcal{PD}_1$ to itself.  The other two transformations act nontrivially, yielding two more phase diagrams
\begin{equation}
    \mathcal{PD}_2 = S_1(\mathcal{PD}_1) = (\text{SPT},\text{SSB}) = O(\mathcal{PD}_4) = S_2(\mathcal{PD}_5) ,
\end{equation}
\begin{equation}
    \mathcal{PD}_3 = S_2(\mathcal{PD}_1) = (\text{Hald},\text{Hald+SSB}) = O(\mathcal{PD}_3) = S_1(\mathcal{PD}_5).
\end{equation}
Nothing acts trivially on $\mathcal{PD}_2$, so we obtain
\begin{equation}
    \mathcal{PD}_4 = O(\mathcal{PD}_2) = (\text{SSB}',\text{Tri}) = S_1(\mathcal{PD}_3) = S_2 (\mathcal{PD}_7)
\end{equation}
and
\begin{equation}
    \mathcal{PD}_5 = (\text{Hald}+\text{SSB},\text{Hald}+\text{SPT}) = S_2(\mathcal{PD}_2) = S_1 (\mathcal{PD}_3) = O(\mathcal{PD}_8).
\end{equation}
We can go further
\begin{equation}
    \mathcal{PD}_6 = (\text{Hald}+\text{SSB}', \text{SPT}) = S_1(\mathcal{PD}_4) = S_2(\mathcal{PD}_8) = O(\mathcal{PD}_8),
\end{equation}
\begin{equation}
    \mathcal{PD}_7 = (\text{Hald}+\text{SSB}',\text{Hald}) = S_2(\mathcal{PD}_4) = O(\mathcal{PD}_5) = S_1(\mathcal{PD}_8),
\end{equation}
and
\begin{equation}
    \mathcal{PD}_8 = (\text{SSB}', \text{Hald}+\text{SPT}) = O(\mathcal{PD}_6) = S_2(\mathcal{PD}_6) = S_1(\mathcal{PD}_7)
\end{equation}

These are summarized in figure \ref{bosonicphasediagrams}.  
\subsection{CFTs}
We now obtain explicit actions for all CFTs that control the phase diagrams in the above subsection.  Since we started with the trivial - SSB transition, we begin with the Ising CFT, written as in \cite{Karch:2019lnn}:
\begin{equation}
    S_1 = \int [(D_A \phi)^2 + \phi^4].
\end{equation}
The invariance of the phase diagram under $O$ is reflected in Kramers-Wannier duality, which allows us to exchange \cite{Karch:2019lnn}
\begin{equation}
    \int [(D_a \phi)^2 + \phi^4 + i\pi a \cup A] \leftrightarrow \int [(D_A \phi)^2 + \phi^4].
\end{equation}
The second phase diagram is obtained by applying $S_1$, which simply maps us to
\begin{equation}
    S_2 = \int [(D_A \phi)^2 + \phi^4 + i\pi A \cup w_1].
\end{equation}
This is a gapless SPT Ising+SPT that governs the transition between SPT and SSB, so we have the correct phase diagram.  Applying $S_2$ to the Ising CFT gives
\begin{equation}
    S_3 = \int [(D_A \phi)^2 + \phi^4 + i\pi w_2],
\end{equation}
which is a gapless SPT Ising+Hald that governs the transition between the Haldene chain and SSB+Haldene.  Applying $O$ to Ising+SPT yields
\begin{equation}
    S_4 = \int [(D_a \phi)^2 + \phi^4 + i\pi a\cup (A+w_1)] = \int [(D_{A+w_1}\phi)^2 + \phi^4].
\end{equation}
This is like the Ising CFT, but the $\mathbb{Z}_2$ background is shifted by $w_1$.  It governs the transition between the trivial and SSB' phases, matching the phase diagram. We'll call it the Ising' CFT.  Applying $S_2$ to Ising + SPT gives
\begin{equation}
    S_5 = \int [(D_A \phi)^2 + \phi^4 + i\pi (A\cup w_1 + w_2)].
\end{equation}
This is the gapless SPT Ising + (SPT+Haldane).  By deforming the mass squared, we can see that it governs the transition between the SPT + Haldane and SSB + Haldane phases, which is the correct phase diagram.  Applying $S_1$ to Ising' gives
\begin{equation}
    S_6 = \int [(D_{A+w_1}\phi)^2 + \phi^4 + i\pi A \cup w_1],
\end{equation}
which is a gapless SPT that we will call Ising'+SPT.  It governs the phase transition between the SPT and SSB' + Hald phases, as it should.  Applying $S_2$ to Ising' gives
\begin{equation}
    S_7 = \int [(D_{A+w_1}\phi)^2 + \phi^4 + i\pi w_2],
\end{equation}
which is a gapless SPT that we will call Ising' + Haldane.  It governs the phase transition between the Haldane and SSB'+Haldane phases, as it should.  Finally, we can apply $S_2$ to Ising' + SPT to obtain:
\begin{equation}
    S_8 = \int [(D_{A+w_1}\phi)^2 + \phi^4 + i\pi A \cup w_1 + i\pi w_2],
\end{equation}
which is a gapless SPT that we will call Ising' + SPT + Haldane.  It governs the expected phase transition - one between SPT + Hald and SSB'.

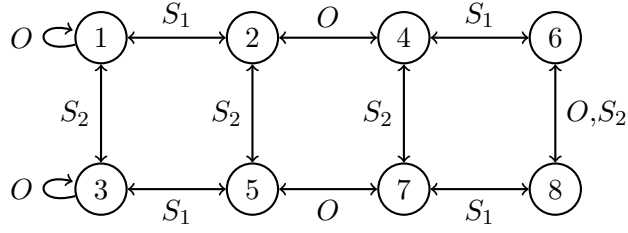
\begin{figure}
\begin{center}
\begin{tikzpicture}[every node,thick] 
 \tikzset{My Style/.style={circle,draw,thick}}

    \node [My Style] (1) at (0,0) {1}; 
    \node [My Style](2) at (2,0) {2}; 
    \node [My Style] (3) at (0,-2) {3};
    \node [My Style](4) at (4,0) {4};
    \node [My Style](5) at (2,-2) {5};
    \node [My Style](6) at (6,0) {6};
    \node [My Style](7) at (4,-2) {7};
    \node [My Style](8) at (6,-2) {8};


    \draw [<->] (1) --   (2) 
    node [midway,above] {$S_1$}; 
    \draw [<->] (2) -- (4)
    node [midway,above] {$O$}; 
    \draw [<->] (4) -- (6)
    node [midway,above] {$S_1$};
    \draw [<->] (1) -- (3)
    node [midway,left] {$S_2$};
    \draw [<->] (2) -- (5)
    node [midway,left] {$S_2$};
    \draw [<->] (3) -- (5)
    node [midway,below] {$S_1$};
    \draw [<->] (5) -- (7)
    node [midway,below] {$O$};
    \draw [<->] (7) -- (8)
    node [midway,below] {$S_1$};
    \draw [<->] (4) -- (7)
     node [midway,left] {$S_2$};
    \draw [<->] (6) -- (8)
    node [midway,right] {$O$,$S_2$};
    \path[<->] (1)
            edge [loop left] node {$O$} ();
    \path[<->](3)
            edge [loop left] node {$O$} ();

\end{tikzpicture}
\caption{Eight CFTs, each one giving rise to a phase transition between two massive phases as described in the text.  They are permuted by the action of $O$, $S_1$, and $S_2$ as depicted. \label{bosonicCFTs}}
\end{center}
\end{figure}
\vskip10pt

\vskip20pt
\begin{figure}
\begin{center}

\begin{tikzpicture}
    \begin{scope}
        \draw[-] (0,-.6) -- (0,.6) ;
        
        \node[font=\scriptsize] at (.4,.4)  {SSB};
        \node[font=\scriptsize] at (-.4,.4) {1};
    \end{scope}
    
 \begin{scope}[xshift=2cm]
        
        \draw[-] (0,-.6) -- (0,.6) ;
        
        \node[font=\scriptsize] at (.4,.4)  {SPT};
        \node[font=\scriptsize] at (-.4,.4) {SSB};
    \end{scope}
    
    \begin{scope}[xshift=4cm]
        \draw[-] (0,-.6) -- (0,.6) ;

        \node[font=\scriptsize] at (.4,.4)  {SSB'};
        \node[font=\scriptsize] at (-.4,.4) {1};
    \end{scope}

     \begin{scope}[xshift=6cm]
        \draw[-] (0,-.6) -- (0,0.6) ;

        \node[font=\scriptsize] at (.6,.4)  {H+SSB'};
        \node[font=\scriptsize] at (-.4,.4) {SPT};
    \end{scope}

     \begin{scope}[xshift=0cm,yshift=-2cm]
        \draw[-] (0,-.6) -- (0,.6) ;
        
        \node[font=\scriptsize] at (.4,.4)  {H};
        \node[font=\scriptsize] at (-.6,.4) {H+SSB};
    \end{scope}

     \begin{scope}[xshift=2cm,yshift=-2cm]
        \draw[-] (0,-.6) -- (0,.6) ;
        
        \node[font=\scriptsize] at (.6,.4)  {H+SSB};
        \node[font=\scriptsize] at (-.6,.4) {H+SPT};
    \end{scope}

     \begin{scope}[xshift=4cm,yshift=-2cm]
        \draw[-] (0,-.6) -- (0,.6) ;
        
        \node[font=\scriptsize] at (.6,.4)  {H+SSB'};
        \node[font=\scriptsize] at (-.4,.4) {H};
    \end{scope}

     \begin{scope}[xshift=6cm,yshift=-2cm]
        \draw[-] (0,-.6) -- (0,.6) ;
        
        \node[font=\scriptsize] at (.4,.4)  {SSB'};
        \node[font=\scriptsize] at (-.4,.4) {H};
    \end{scope}
    
   \draw[<->] (.8,0) -- (1.2,0) ;
    \draw[<->] (2.8,0) -- (3.2,0) ;
     \draw[<->] (4.8,0) -- (5.2,0) ;
     \draw[<->] (.8,-2) -- (1.2,-2);
     \draw[<->] (2.8,-2) -- (3.2,-2);
     \draw[<->] (4.8,-2) -- (5.2,-2);
     \draw[<->] (0,-.7) -- (0,-1.2);
     \draw[<->] (2,-.7) -- (2,-1.2);
     \draw[<->] (4,-.7) -- (4,-1.2);
     \draw[<->] (6,-.7) -- (6,-1.2);

\end{tikzpicture}
\caption{Phase diagrams for the CFTs from Figure \ref{bosonicCFTs}. To avoid clutter we further abbreviated Tri to 1 and Hald to H. \label{bosonicphasediagrams}}
\end{center}
\end{figure}
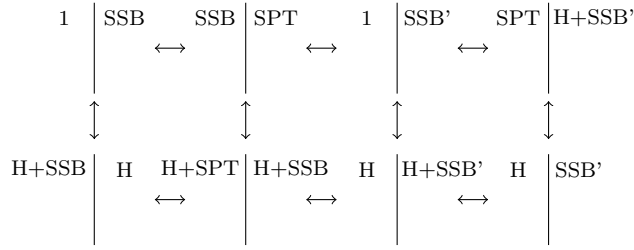
\vskip10pt

\section{Fermionic Web}
In this section we discuss phase transitions between different gapped phases with $\mathbb{Z}_2^F \times \mathbb{Z}_2^T$ symmetry.  We begin by introducing $\text{Pin}^-$ structures and the accompanying Arf-Brown-Kervaire (ABK) invariant that are generally required to define fermions on manifolds that do not admit a Spin structure.  We specify how the ABK invariant is used to fermionize bosonic phases.  By fermionizing bosonic phases, we obtain eight gapped fermionic phases.  We rederive their $\mathbb{Z}_8$ stacking rule discussed in i.e. \cite{Fidkowski:2010jmn,Thorngren:2018bhj}.  With all of that in place we obtain eight phase diagrams, which turn out to be the eight that one would obtain by fermionizing the eight bosonic phase diagrams and address the delicate art of turning a circle into a ladder, which is key to making contact with the web of bosonic phase diagrams.  We then repeat this process for the CFTs controlling the phase diagrams.
\subsection{Spin, Pin, Time-Reversal, Arf, Brown, and Kervaire}
If we wish for our fermions to have time reversal symmetry, a spin structure is no longer the correct tangential structure to require.  Instead, the correct tangential structure is a $\text{Pin}^-$ structure $\eta$ \footnote{Let's be a little more detailed.  A spin structure on an orientable d-manifold is a double cover of the tangent bundle $SO(d)$. Sections of the resulting bundle are called spinors.  On non-orientable manifolds one instead double covers $O(d)$, resulting a pin structure.  The sections of the resulting bundle are called pinors.  There are actually two choices, depending on whether reflections square to positive or negative one.  The $\text{Pin}^-$ structure corresponds to the latter option.  We consider it in this paper since it ``always available" on closed 2-manifolds in the sense that its obstruction, which is $w_1 \cup w_1 + w_2$, always vanishes in that setting, as discussed when we introduced Stiefel-Whitney classes.  This turns out to describe fermions with time reversal symmetry $T^2 = 1$.}.  Correspondingly, the correct quadratic refinement of the intersection pairing is no longer the Arf invariant, but a $\mathbb{Z}_8$-valued invariant called the Arf-Brown-Kevaire (ABK) invariant.  We can define it using:
\begin{equation}
    Z_{\text{ABK}}[\eta] = \exp[\frac{\pi i}{4}\text{ABK}(\eta)] = \sum_{a \in H^1(X;\mathbb{Z}_2)} \exp[\frac{\pi}{2}iq_\eta(a)],
\end{equation}
where $q: H^1(X;\mathbb{Z}_2) \rightarrow \mathbb{Z}_4$ satisfies \footnote{The last term is to be read as an inclusion of $\int a \cup b$ from $\mathbb{Z}_2$ to $\mathbb{Z}_4$ rather than as $0$.}
\begin{equation}
    q_\eta(a+b) = q_\eta(a)+q_\eta(b)+2\int a\cup b.
\end{equation}
Intuitively, it may be viewed as twice of the quadratic form that defines the Arf invariant.  This occurs when $w_1 = 0$ and the manifold is orientable.  $q$ depends on the $\text{Pin}^-$ structure:
\begin{equation}
    q_{B\cdot\eta}(a) = q_\eta(a) + 2 \int a \cup B.
\end{equation}
This allows us to write $Z_{\text{ABK}}$ coupled to a $\mathbb{Z}_2$ background:
\begin{equation}
    Z_{\text{ABK}}[A] = \exp[\frac{\pi i}{4}\text{ABK}(A\cdot \eta)] = \sum_{a \in H^1(X;\mathbb{Z}_2)} \exp[\frac{\pi i}{2}q_\eta (a) + \pi i \int a \cup A].
\end{equation}
Indeed, we can go further and imitate \cite{Gaiotto:2020iye} by using 
\begin{equation}
    q_\eta (a) = \frac{1}{2}[\text{ABK}(a\cdot \eta) - \text{ABK}(\eta)] 
\end{equation}
so that
\begin{equation}
    \exp[\frac{\pi i}{4}\text{ABK}(A\cdot \eta)] = \sum_{a \in H^1(X;\mathbb{Z}_2)} \exp[\frac{\pi i}{4}(\text{ABK}(a\cdot \eta)-\text{ABK}(\eta)) + \pi i\int a\cup A].
\end{equation}
Moreover, one can use the quadratic property of $q_\eta (A)$ to show
\begin{equation}
    \text{ABK}((A+B)\cdot \eta) = \text{ABK}(A\cdot \eta) + \text{ABK}(B\cdot \eta) - \text{ABK}(\eta) + 4\int A \cup B.
\end{equation}
This formula and the above will be critical \footnote{That is to say ``of utmost importance" rather than ``located at a phase transition."} in the following.  We will use the ABK invariant to fermionize a bosonic theory with $\mathbb{Z}_2$ symmetry according to the formula:
\begin{equation}
    Z_F[C] = \sum_{a \in H^1(X;\mathbb{Z}_2)} Z_B[a] \exp[\frac{\pi i}{4}(\text{ABK}(a\cdot \eta)-\text{ABK}(\eta)) + \pi i\int a\cup C] .
\end{equation}
The inverse of this map is bosonization.  It is generally given by a sum over $\text{Pin}^-$ structures
\begin{equation}
    Z_B[A] = \sum_\eta \exp[\frac{\pi i}{4}(\text{ABK}(A\cdot \eta)-\text{ABK}(\eta))+\pi i\int A \cup w_1]Z_F[\eta],
\end{equation}
where, for clarity, we have restored the dependence of the fermionic partition function on the $\text{Pin}^-$ structure.  Since $\text{Pin}^-$ structures form an $H^1(X;\mathbb{Z}_2)$ torsor, summing over gauge fields will also include a sum over $\text{Pin}^-$ structures.  We can gauge $C$ in the fermionic theory
\begin{equation}
    Z_{F'}[A] = \sum_{c \in H^1(X;\mathbb{Z}_2)} Z[c\cdot \eta] (-1)^{\int c \cup A}.
\end{equation}
Using the quadratic property of the ABK invariant, we can use the above to write 
\begin{equation}
    Z_B[A] = \sum_{c \in H^1(X;\mathbb{Z}_2)} Z_F[c] \exp[-\frac{\pi i}{4}(\text{ABK}(A\cdot \eta)-\text{ABK}(\eta)) + \pi i \int c \cup A],
\end{equation}
which describes the bosonization of a fermionic theory by gauging fermion parity.  While it will not concern us in this paper, it is worth noting that gravitational anomalies generally prevent one from bosonizing by gauging fermion parity \cite{BoyleSmith:2024qgx}.

\subsection{Gapped Phases and Topological Operations}
We can determine the gapped fermionic phases by fermionizing the bosonic ones.  By definition, fermionizing $Z_{\text{Tri}}[A]$ gives $Z_{\text{ABK}}[C]$.  Moreover, it is obvious that fermionizing $Z_{\text{SSB}}[A]$ gives $Z_{\text{Tri}}[C]$.  Fermionizing $Z_{\text{SPT}}[A]$ gives:
\begin{multline}
    \sum_{a \in H^1(X;\mathbb{Z}_2)} \exp[\frac{\pi i}{4}(\text{ABK}(a\cdot \eta)-\text{ABK}(\eta)) + \pi i\int a\cup (C+w_1)] \\= \exp[\frac{\pi i}{4}\text{ABK}(C+w_1)]= Z_{\text{ABK}+\text{SPT}+\text{Hald}} [C].
\end{multline}
We will justify the name of the phase later when we discuss stacking with the ABK invertible theory.

Fermionizing $Z_{\text{Hald}}$ gives
\begin{equation}
    Z_{\text{ABK+Hald}}[C] = \exp[\frac{\pi i}{4} \text{ABK}(C\cdot \eta) + \pi i \int w_2].
\end{equation}
Fermionizing $Z_{\text{SSB'}}[A]$ gives
\begin{multline}
    \sum_{a \in H^1(X;\mathbb{Z}_2)} \delta(a+w_1) \exp[\frac{\pi i}{4}(\text{ABK}(a\cdot \eta)-\text{ABK}(\eta)) + \pi i\int a\cup C] \\= \exp[\frac{\pi i}{4}(\text{ABK}(w_1\cdot \eta)-\text{ABK}(\eta)) + \pi i\int w_1\cup C] = Z_{\text{SPT}} [C,w_1].
\end{multline}
Fermionizing $Z_{\text{Hald+SSB}}$ gives
\begin{equation}
    Z_{\text{Hald}} = (-1)^{\int w_2},
\end{equation}
which is just as it was in the bosonic case.  Indeed, this theory and the trivial theory are actually bosonic theories, since they do not depend on the $\text{Pin}^-$ structure of the spacetime.  Fermionizing $Z_{\text{Hald}+\text{SPT}}$ gives
\begin{equation}
    Z_{\text{ABK}+\text{SPT}}[C] = \exp[\frac{\pi i}{4}\text{ABK}(C+w_1) + \pi i \int w_2].
\end{equation}
Fermionizing $Z_{\text{Hald + SSB'}}$ gives
\begin{equation}
    Z_{\text{SPT} + \text{Hald}}[C] =  \exp[\frac{\pi i}{4}(\text{ABK}(w_1\cdot \eta)-\text{ABK}(\eta)) + \pi i(\int w_1\cup C+\int w_2)]. 
\end{equation}
The correspondence between fermionic and bosonic phases, as well as the correspondence between bosonic and fermionic topological operations that we will discuss later are summarized in table \ref{fermboscorr}.

\begin{table}[]
    \centering
    \begin{tabular}{|c|c|}
    \hline
       Bosonic  & Fermionic \\
    \hline
       Tri  & ABK \\
    \hline
        SSB & Tri \\
    \hline 
        SPT & ABK+SPT+Hald \\
    \hline 
        SSB' & SPT \\
    \hline 
        Hald & Hald + ABK \\
    \hline 
        Hald + SSB & Hald \\
    \hline
        Hald + SPT & ABK+SPT \\
    \hline 
        Hald + SSB' & SPT + Hald \\
    \hline 
        $O$ & $O_F$ \\
    \hline 
        $S_1$ & $S_1^F$ \\
    \hline 
        $S_2$ & $S_2^F$ \\
    \hline 
    \end{tabular}
    \caption{The correspondence between bosonic phases and the fermionic phases obtained by fermionizing them, as well as the correspondence between bosonic and fermionic topological operations.  As discussed in the main text, $O$ is gauging $\mathbb{Z}_2$, $O_F$ first shifts fermion parity by $w_1$ then stacks ABK, $S_1$ stacks SPT, $S_1^F$ shifts fermioin parity by $w_1$, $S_2$ stacks the Haldane chain and $S_2^F$ does the same.  We have not seen the correspondses between the bosonic and fermionic operations stressed in the literature.}
    \label{fermboscorr}
\end{table}

\cite{Fidkowski:2010jmn} showed that invertible phases of time reversal invariant fermions in 2 dimensions obey a $\mathbb{Z}_8$ classification.  Let us show that our phases stack accordingly.  We will index them by $\nu \in \mathbb{Z}_8$.  Clearly, the trivial phase is $\nu = 0$ and the ABK phase is $\nu = 1$.  Stacking ABK on itself yields
\begin{multline}
    \sum_{a,a'} \exp[\frac{\pi i}{2}(q_\eta(a)+q_\eta(a'))+\pi i \int (a+a')\cup C] \\= \sum_{a,a'} \exp[\frac{\pi i}{2}q_\eta(a+a') + \pi i\int a \cup a' + \pi i\int (a+a')\cup C ] \\ = \sum_{a'',a} \exp[\frac{\pi i}{2}q_\eta(a'') + \pi i \int a''\cup C +\pi i\int a \cup (a''+a)] \\= \sum_{a'',a} \exp[\frac{\pi i}{2}q_\eta(a'') + \pi i \int a''\cup C +\pi i\int a \cup (a''+w_1)] \\ = \exp[\frac{\pi i}{2}q_\eta(w_1) + \pi i \int w_1 \cup C] \\= \exp[\frac{\pi i}{4}(\text{ABK}(w_1\cdot \eta)-\text{ABK}(\eta)) + \pi i\int w_1 \cup C] = Z_{\text{SPT}}[C,w_1].
\end{multline}
The first equality comes from the quadratic property of $q_\eta$.  The second equality comes from changing variables in the integral.  The third equality uses $a \cup a = a \cup w_1$.  The fourth equality comes from integrating out $a$, and the fifth equality makes precise our choice of quadratic function.  $\nu = 3$ corresponds to:
\begin{multline}
    \sum_{a,b,c} \exp[\frac{\pi i}{2}(q_\eta(a) + q_\eta(b) + q_\eta(c)) + \pi i \int (a+b+c)\cup C] \\ = \sum_{a,b,c}\exp[\frac{\pi i}{2}(q_\eta(a+b)+q_\eta(c)) + \pi i \int a \cup b + \pi i\int (a+b+c)\cup C] \\ = \sum_{a',b,c} \exp[\frac{\pi i}{2}(q_\eta(a') + q_\eta(c)) + \pi i\int (a'+b)\cup b + \pi i\int (a'+c)\cup C] \\ = \sum_{a',b,c} \exp[\frac{\pi i}{2}(q_\eta(a') + q_\eta(c)) + \pi i\int a'\cup b + \pi i \int w_1\cup b + \pi i\int (a'+c)\cup C] \\ = \sum_c \exp[\frac{\pi i}{2}(q_\eta(w_1)+q_\eta(c)) + \pi i\int (w_1+c)\cup C] \\ = \sum_c \exp[\frac{\pi i}{2}q_\eta(w_1 +c) + \pi i \int w_1 \cup c + \pi i \int (w_1 +c)\cup C] \\ = \sum_{c'} \exp [\frac{\pi i}{2}q_\eta(c') + \pi i \int (c'+w_1)\cup w_1 + \pi i\int c' \cup C] \\ = \exp[\frac{\pi i}{4}\text{ABK}((w_1+C)\cdot \eta) + \pi i \int w_2] = Z_{\text{ABK}+\text{SPT}}
\end{multline}
Reviewing the calculation also shows that 
\begin{equation}
    Z_{\text{ABK}+\text{SPT}} = Z_{\text{ABK}}[C]Z_{SPT},
\end{equation}
which is how it is written in \cite{Thorngren:2018bhj} and clearly justifies our titling of the phase.  $\nu =4$ corresponds to:
\begin{multline}
    \sum_{a,b,c,d} \exp[\frac{\pi i}{2}(q_\eta(a)+q_\eta(b)+q_\eta(c)+q_\eta(d)) + \pi i \int (a+b+c+d)\cup C] \\ = \sum_{a,b,c,d}\exp[\frac{\pi i}{2}(q_\eta(a+b)+q_\eta(c)+q_\eta(d)) + \pi i \int a \cup b +\pi i\int (a+b+c+d)\cup C]\\ = \sum_{a,b,c,d}\exp[\frac{\pi i}{2}(q_\eta(a+b+c)+q_\eta(d)) + \pi i \int( (a+b)\cup c + a\cup b + (a+b+c+d)\cup C)] \\ =\sum_{a,b,c,d}\exp[\frac{\pi i}{2}q_\eta(a+b+c+d) + \pi i \int((a+b+c)\cup d +(a+b)\cup c + a\cup b + (a+b+c+d)\cup C)] \\ = \sum_{a',b,c,d} \exp [\frac{\pi i}{2}q_\eta(a')+ \pi i \int ((a'+d)\cup d + (a'+c+d)\cup c + (a'+b+c+d)\cup b + a'\cup C)]\\ = \sum_{a',b,c,d} \exp [\frac{\pi i}{2}q_\eta(a')+ \pi i \int ((a'+w_1)\cup d + (a'+w_1+d)\cup c + (a'+w_1+c+d)\cup b + a'\cup C)] \\ = \sum_{a',d} \exp [\frac{\pi i}{2}q_\eta(a')+ \pi i \int ((a'+w_1)\cup d + (a'+w_1+d)\cup (a'+w_1+d) + a'\cup C)] \\ = \sum_{a',d} \exp [\frac{\pi i}{2}q_\eta(a')+ \pi i \int (a'\cup d + (a'+w_1)\cup w_1 + a'\cup C)] \\ = (-1)^{\int w_2} = Z_{\text{Hald}},
\end{multline}
where the final equality comes from integrating out $d$, noting that $q_\eta(0)=0$, and then noting that $w_1\cup w_1 = w_2$.  Clearly, $\nu = 5$ corresponds to ABK+Haldane, $\nu = 6$ corresponds to 5+Haldane, and $\nu =7$ corresponds to ABK+SPT+Hald.  This recreation of the result in \cite{Fidkowski:2010jmn} from a field theory perspective also appears in \cite{Thorngren:2018bhj}.  We summarize it in the following table:
\begin{table}[]
    \centering
    \begin{tabular}{|c|c|}
        \hline
        $\nu = 0$ & $Z_{\text{Tri}}[C] = 1$ \\
        \hline
        $\nu = 1$ & $Z_{\text{ABK}}[C] = \exp[\frac{\pi i}{4}\text{ABK}(C\cdot\eta)]$ \\
        \hline
        $\nu = 2$ & $Z_\text{SPT}[C,w_1] = \exp[\frac{\pi i}{4}(\text{ABK}(w_1\cdot \eta)-\text{ABK}(\eta))+\pi i\int w_1 \cup C]$ \\
        \hline
        $\nu = 3$ & $Z_{\text{ABK}+\text{SPT}}[C] = \exp [\frac{\pi i}{4}\text{ABK}((C+w_1)\cdot\eta)+\pi i \int w_2)]$ \\
        \hline
        $\nu = 4$ & $Z_{\text{Hald}}[C] = (-1)^{\int w_2}$ \\
        \hline
        $\nu = 5$ & $Z_{\text{ABK+Hald}}[C] = \exp[\frac{\pi i}{4}\text{ABK}(C\cdot\eta)+i\pi \int w_2]$\\
        \hline
        $\nu = 6$ & $Z_{\text{SPT} + \text{Hald}}[C] = \exp[\frac{\pi i}{4}(\text{ABK}(w_1\cdot \eta)-\text{ABK}(\eta))+\pi i\int w_1 \cup C + \pi i \int w_2]$\\
        \hline
        $\nu = 7$ & $Z_{\text{ABK}+\text{SPT}+\text{Hald}}[C] = \exp [\frac{\pi i}{4}\text{ABK}((C+w_1)\cdot\eta)]$\\
        \hline
    \end{tabular}
    \caption{The $\mathbb{Z}_8$ stacking of phases with $\mathbb{Z}_2^F \times \mathbb{Z}_2^T$ symmetry.  $\nu=0$, $\nu=2$, $\nu=4$, and $\nu=6$ correspond to SPT phases, in the sense that they are trivial if the backgrounds for fermion parity and time-reversal symmetry are turned off.  These obey a $\mathbb{Z}_4$ classification.}
    \label{tab:my_label}
\end{table}

Clearly, stacking with the appropriate power of ABK maps us between any two fermionic phases. However, the stacking operation lacks a clear bosonic analog \footnote{Indeed, its analog looks more like a Fourier transform than any of our operations, see \cite{Thorngren:2018bhj}.}.  To better understand the relationship between the bosonic webs and the fermionic ones that arise by fermionizing them, we seek more direct analogs to the bosonic operations.  Obviously, the analog of stacking with the Haldane phase is stacking with the Haldane phase.  Stacking with the SPT shifts the fermion parity by $w_1$.  The most subtle is gauging $\mathbb{Z}_2$, which corresponds to shifting the the fermion parity by the time-reversal symmetry and then stacking ABK.  The actions of these various operations, which we call $S_1^F$, $S_2^F$, and $O_F$, respectively, act on the various gapped phases as indicated in table \ref{fermorbgrp}.

\begin{table}[]
    \centering
    \begin{adjustbox}{width=\linewidth}
    \begin{tabular}{|c|c|c|c|c|c|c|c|c|}
    \hline
    $\mathcal{T}$ & Tri & ABK & SPT & ABK+SPT & Hald & ABK+Hald & SPT+Hald & ABK+SPT+Hald \\
    \hline
    $O_F(\mathcal{T})$ & ABK & Tri & ABK+SPT+Hald  & SPT+Hald & ABK+Hald & Hald & ABK+SPT & SPT \\
    \hline
    $S_1^F(\mathcal{T})$ & Tri & ABK+SPT+Hald & SPT+Hald & ABK+Hald & Hald  & ABK+SPT & SPT & ABK \\ 
    \hline
    $S_2^F(\mathcal{T})$ & Hald & ABK+Hald & SPT+Hald & ABK+SPT+Hald & Triv & ABK & SPT & ABK+SPT \\
    \hline
    \end{tabular}
    \end{adjustbox}
    \caption{The action of the topological manipulations $O_F$ (shifting $C$ by $w_1$ then stacking ABK), $S_1$ (shifting $C$ to $C+w_1$), and $S_2$ (stacking the Haldene chain) on a gapped theory $\mathcal{T}$.  The various choices of $\mathcal{T}$ are as described in the text.  These operations map $\nu$ to $-\nu +1$, $-\nu$, and $\nu+4$, respectively.}
    \label{fermorbgrp}
\end{table}
\subsection{Phase Diagrams}
We now discuss the phase diagrams.  Our strategy is familiar, we begin with a seed phase diagram and generate others by either stacking ABK or shifting the background by the first Stiefel-Whitney class.  Our seed is the trivial to ABK transition:
\begin{equation}
    \mathcal{PD}_1 = (\text{Tri},\text{ABK}).
\end{equation}
Stacking repeatedly generates:
\begin{equation}
    \mathcal{PD}_2 = (\text{ABK},\text{SPT}),
\end{equation}
\begin{equation}
    \mathcal{PD}_3 = (\text{SPT},\text{SPT}+\text{ABK}),
\end{equation}
\begin{equation}
    \mathcal{PD}_4 = (\text{ABK}+\text{SPT},\text{Hald}),
\end{equation}
\begin{equation}
    \mathcal{PD}_5 = (\text{Hald},\text{ABK+Hald}) ,
\end{equation}
\begin{equation}
    \mathcal{PD}_6 = (\text{ABK+Hald},\text{SPT}+\text{Hald}),
\end{equation}
\begin{equation}
    \mathcal{PD}_7 = (\text{SPT}+\text{Hald},\text{ABK}+\text{SPT}+\text{Hald}) ,
\end{equation}
and
\begin{equation}
    \mathcal{PD}_8 = (\text{ABK}+\text{SPT}+\text{Hald},\text{Tri}) .
\end{equation}
Of course, the way these phase diagrams are obtained looks nothing like the bosonic process.  The remedy to this is in the fact that the operations in table \ref{fermorbgrp} act on the phase diagrams in a way analogous to their bosonic counterparts.  The results are in figure \ref{fermphasediagrams}.

\subsection{CFTs}
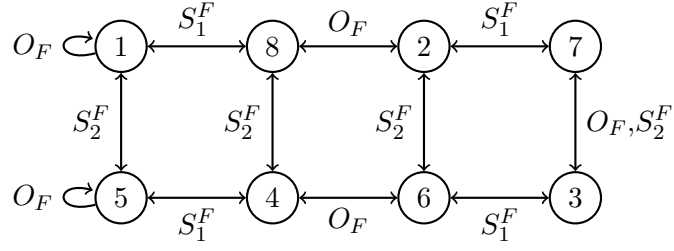
\begin{figure}
\begin{center}
\begin{tikzpicture}[every node,thick] 
 \tikzset{My Style/.style={circle,draw,thick}}

    \node [My Style] (1) at (0,0) {1}; 
    \node [My Style](8) at (2,0) {8}; 
    \node [My Style] (5) at (0,-2) {5};
    \node [My Style](2) at (4,0) {2};
    \node [My Style](4) at (2,-2) {4};
    \node [My Style](7) at (6,0) {7};
    \node [My Style](6) at (4,-2) {6};
    \node [My Style](3) at (6,-2) {3};


    \draw [<->] (1) --   (8) 
    node [midway,above] {$S_1^F$}; 
    \draw [<->] (8) -- (2)
    node [midway,above] {$O_F$}; 
    \draw [<->] (2) -- (7)
    node [midway,above] {$S_1^F$};
    \draw [<->] (1) -- (5)
    node [midway,left] {$S_2^F$};
    \draw [<->] (8) -- (4)
    node [midway,left] {$S_2^F$};
    \draw [<->] (4) -- (5)
    node [midway,below] {$S_1^F$};
    \draw [<->] (4) -- (6)
    node [midway,below] {$O_F$};
    \draw [<->] (6) -- (3)
    node [midway,below] {$S_1^F$};
    \draw [<->] (2) -- (6)
     node [midway,left] {$S_2^F$};
    \draw [<->] (7) -- (3)
    node [midway,right] {$O_F$,$S_2^F$};
    \path[<->] (1)
            edge [loop left] node {$O_F$} ();
    \path[<->](5)
            edge [loop left] node {$O_F$} ();

\end{tikzpicture}
\caption{Eight CFTs, each one giving rise to a phase transition between two massive phases as described in the text.  They are permuted by the action of $O$, $S_1^F$, and $S_2^F$ as depicted. \label{fermCFTs}}
\end{center}
\end{figure}

\begin{figure}
\begin{center}

\begin{tikzpicture}
    \begin{scope}
        \draw[-] (0,-.6) -- (0,.6) ;
        
        \node[font=\scriptsize] at (.4,.4)  {ABK};
        \node[font=\scriptsize] at (-.4,.4) {1};
    \end{scope}
    
 \begin{scope}[xshift=2cm]
        
        \draw[-] (0,-.6) -- (0,.6) ;
        
        \node[font=\scriptsize] at (.4,.4)  {ASH};
        \node[font=\scriptsize] at (-.4,.4) {1};
    \end{scope}
    
    \begin{scope}[xshift=4cm]
        \draw[-] (0,-.6) -- (0,.6) ;

        \node[font=\scriptsize] at (.4,.4)  {SPT};
        \node[font=\scriptsize] at (-.4,.4) {ABK};
    \end{scope}

     \begin{scope}[xshift=6cm]
        \draw[-] (0,-.6) -- (0,0.6) ;

        \node[font=\scriptsize] at (.4,.4)  {HS};
        \node[font=\scriptsize] at (-.4,.4) {ASH};
    \end{scope}

     \begin{scope}[xshift=0cm,yshift=-2cm]
        \draw[-] (0,-.6) -- (0,.6) ;
        
        \node[font=\scriptsize] at (.6,.4)  {H+ABK};
        \node[font=\scriptsize] at (-.4,.4) {H};
    \end{scope}

     \begin{scope}[xshift=2cm,yshift=-2cm]
        \draw[-] (0,-.6) -- (0,.6) ;
        
        \node[font=\scriptsize] at (.3,.4)  {H};
        \node[font=\scriptsize] at (-.4,.4) {AS};
    \end{scope}

     \begin{scope}[xshift=4cm,yshift=-2cm]
        \draw[-] (0,-.6) -- (0,.6) ;
        
        \node[font=\scriptsize] at (.4,.4)  {HS};
        \node[font=\scriptsize] at (-.6,.4) {H+ABK};
    \end{scope}

     \begin{scope}[xshift=6cm,yshift=-2cm]
        \draw[-] (0,-.6) -- (0,.6) ;
        
        \node[font=\scriptsize] at (.4,.4)  {SPT};
        \node[font=\scriptsize] at (-.4,.4) {AS};
    \end{scope}
    
   \draw[<->] (.8,0) -- (1.2,0) ;
    \draw[<->] (2.8,0) -- (3.2,0) ;
     \draw[<->] (4.8,0) -- (5.2,0) ;
     \draw[<->] (.8,-2) -- (1.2,-2);
     \draw[<->] (2.8,-2) -- (3.2,-2);
     \draw[<->] (4.8,-2) -- (5.2,-2);
     \draw[<->] (0,-.7) -- (0,-1.2);
     \draw[<->] (2,-.7) -- (2,-1.2);
     \draw[<->] (4,-.7) -- (4,-1.2);
     \draw[<->] (6,-.7) -- (6,-1.2);

\end{tikzpicture}
\caption{Phase diagrams for the CFTs from Figure \ref{fermCFTs}. To avoid clutter we further abbreviated Tri to 1, Hald to H, ABK+SPT+Hald to ASH, Hald+SPT to HS, and ABK+SPT to AS. \label{fermphasediagrams}}
\end{center}
\end{figure}
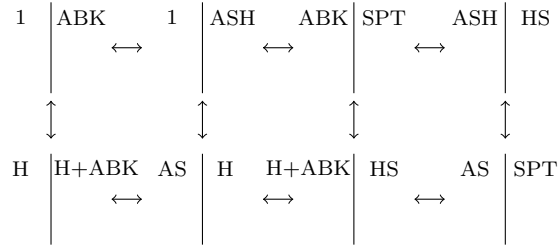

We claim that the seed CFT is simply the Majorana CFT coupled to a $\text{Pin}^-$ structure rather than a Spin structure.  Its action is
\begin{equation}
    S_1 = \int i\bar{\chi} \slashed{D}_{C\cdot\eta} \chi. 
\end{equation}
To see that this is the case, note that the ABK invariant is an $\eta$ invariant that emerges from the regularization of massive, free Majorana fermions.  In particular
\begin{equation}
    \exp[\frac{\pi i}{4}\text{ABK}(C\cdot \eta)] = \frac{Z_{Maj}[C,m>>0]}{Z_{Maj}[C,m<<0]}.
\end{equation}
Thus, using the only relevant deformation of the above CFT, we see that it controls the phase transition between the trivial and ABK phases.  With our seed CFT in place, we are free to apply the isomorphisms of our orbifold groupoid to obtain the CFTs governing the other phase transitions.  As discussed above, this is done simply by stacking ABK over and over.  The results are:
\begin{equation}
    S_2 = \int [i\bar{\chi}\slashed{D}_{C\cdot \eta}\chi + \frac{\pi i}{4}\text{ABK}(C\cdot \eta) ],
\end{equation}
\begin{equation}
    S_3 = \int [i\bar{\chi}\slashed{D}_{C\cdot \eta}\chi] + \frac{\pi i}{4}[\text{ABK}(w_1 \cdot \eta)- \text{ABK}(\eta)] + i\pi \int w_1 \cup C,
\end{equation}
\begin{equation}
    S_4 = \int [i\bar{\chi}\slashed{D}_{C\cdot \eta}\chi] + \frac{\pi i}{4}\text{ABK}((C+w_1)\cdot \eta) + \pi i \int w_2,
\end{equation}
\begin{equation}
    S_5 = \int [i\bar{\chi}\slashed{D}_{C\cdot \eta} \chi] + \pi i \int w_2,
\end{equation}
\begin{equation}
    S_6 = \int [i\bar{\chi}\slashed{D}_{C\cdot \eta}\chi] + \frac{\pi i}{4}\text{ABK}(C \cdot \eta) + \pi i\int w_2
\end{equation}
\begin{equation}
    S_7 = \int [i\bar{\chi}\slashed{D}_{C\cdot \eta}\chi] + \frac{\pi i}{4}[\text{ABK}(w_1 \cdot \eta) - \text{ABK}(\eta)] + \pi i \int[w_1 \cup C + w_2],
\end{equation}
and
\begin{equation}
    S_8 = \int [i\bar{\chi}\slashed{D}_{C\cdot\eta}\chi] + \frac{\pi i}{4}\text{ABK}((C+w_1) \cdot \eta).
\end{equation}
Let us pause to make some comments:
\begin{itemize}
    \item These all have the correct phase diagrams, since their phase diagrams are simply obtained by stacking the relevant gapped phase on top of $\mathcal{PD}_1$.
    \item $S_2$ simply stacks Majorana and ABK.  It is the analog of the Majorana + Arf partially gapped phase in \cite{Karch:2025jem}.
    \item $S_3$, $S_5$, and $S_7$ are all gapless SPT phases that stack one of the nontrivial SPTs on top of the Majorana CFT.  
    \item $S_4$, $S_6$, and $S_8$ are of a similar mold - they clearly stack an invertible theory on top of the Majorana CFT.  We do not call them gapless SPTs since the invertible theory in question does not become trivial when one turns of the backgrounds for the $\mathbb{Z}_2$ and time-reversal symmetries.
\end{itemize}
These CFTs were all obtained by directly stacking the ABK phase multiple times.  However, as illustrated in figure \ref{fermCFTs}, this is not the only way to move between CFTs.  Let us weaponize this to discover some dualities.  First, note that to get from $S_1$ to $S_8$, we can apply $S_1^F$.  To make this consistent, we see that we require:
\begin{equation}
    \int i\bar{\chi}\slashed{D}_{(C+w_1)\cdot \eta}\chi \leftrightarrow \int [i\bar{\chi}\slashed{D}_{C\cdot\eta}\chi] + \frac{\pi i}{4}\text{ABK}((C+w_1) \cdot \eta),
\end{equation}
that is shifting the $\text{Pin}^-$ structure by the time reversal symmetry amounts to stacking ABK+SPT+Hald.  Moreover, the fact that shifting the fermion parity and stacking ABK leaves $S_1$ invariant leads to the requirement that 
\begin{equation}
    \int [i\bar{\chi}\slashed{D}_{(C+w_1)\cdot \eta}\chi + \frac{\pi i}{4}\text{ABK}(C\cdot \eta)] \leftrightarrow \int [i\bar{\chi}\slashed{D}_{C\cdot \eta}\chi],
\end{equation}
which, upon recalling that $Z_{\text{ABK}+\text{SPT}+\text{Hald}}$ is the inverse of $Z_{\text{ABK}}$, actually follows from the above duality.  This is the fermionic analog of Kramers-Wannier duality.  As a sanity check, if the manifold is orientable then $w_1 = 0$ and the above restricts to the fact that stacking the Arf theory on the Majorana CFT yields the Majorana CFT.  One can repeat this throughout the figure to find dualities whose veracity depends on the first.    

\vskip10pt

\vskip20pt
\vskip10pt

\subsection{Comments on $\text{Pin}^+$ Theories}
So far, we have discussed time-reversal symmetric fermion theories with $T^2=1$.  These couple to a $\text{Pin}^-$ structure.  We now make some comments on time-reversal symmeric fermionic theories with $T^2 = (-1)^F$.  These theories couple to a $\text{Pin}^+$ structure.  Unlike Spin and $\text{Pin}^-$ structures, $\text{Pin}^+$ structures do not correspond in a natural way to quadratic refinements of the intersection pairing, so our analysis of gapped phases that we used to obtain phase transitions does not directly carry over to this case.  Fortunately, as discussed in \cite{Kaidi:2019tyf}, $\text{Pin}^+$ structures on a non-orientable manifold $X$ are in one to one correspondence with Spin structures on its oriented double cover $\tilde{X}$, so we can say something.  Thus, there is a single nontrivial invertible theory coupled to a $\text{Pin}^+$ structure.  Its action is
\begin{equation}
    S = i\pi \text{Arf}[C\cdot \rho],
\end{equation}
where $\rho$ is a Spin structure on $\tilde{X}$.  The CFT mediating the transition between this theory and the trivial theory is the Majorana CFT on $\tilde{X}$.  As discussed in \cite{Witten:2015aba}, when mapping the Majorana CFT back to $X$, we have two choices of $\text{Pin}^+$ structure, related by action of $w_1$.  It would be interesting to examine the impact of this on the phase diagrams in future work - since $\text{Pin}^+$ structures do not correspond to quadratic refinements, doing so would involve methodology not discussed in this paper.

\section{Discussion}
In this paper, we used topological manipulations such as gauging finite symmetries and stacking invertible phases to study phase transitions between time-reversal symmetric theories.  We began with a well understood seed CFT and then mapped it to CFTs describing less familiar phase transitions.  We examined transitions between bosonic theories and between fermionic theories.  On the bosonic side, we reviewed the role of $w_1$ as a background field for time reversal symmetry.  We discussed the gapped phases with $\mathbb{Z}_2 \times \mathbb{Z}_2^T$ symmetry and the maps between them.  Using this, we obtained simple descriptions of many phase transitions, including transitions between phases that break the $\mathbb{Z}_2$ symmetry and SPT phases protected by either time reversal or time-reversal and $\mathbb{Z}_2$.  Many of the CFTs we uncover are gapless SPT phases.  One the fermionic side, we reviewed the connection between quadratic refinements, $\text{Pin}^-$ structures, and the $\mathbb{Z}_8$ classification of fermionic theories with $\mathbb{Z}_2^F \times \mathbb{Z}_2^T$ symmetry in the case where $T^2 = 1$.  We discussed the maps between these gapped phases and the correspondence between these maps and their bosonic counterparts.  We then applied all of this to obtain the CFTs that govern transitions between the fermionic phases that exist within the $\mathbb{Z}_8$ classification in \cite{Fidkowski:2010jmn}.  We also made some comments about theories that depend on $\text{Pin}^+$ structures.  Many of these CFTs are the Majorana CFT stacked with an invertible field theory.  For these theories, $T^2 = (-1)^F$.  We conclude by highlighting some interesting avenues for further research:
\begin{itemize}
    \item As discussed in the body of the paper, the fact that $\text{Pin}^+$ structures do not correspond to quadratic refinements means that our methodology cannot determine the impact of shifting the $\text{Pin}^+$ structure by $w_1$ on the phase transitions.  This merits analysis elsewhere.
    \item It would be worthwhile to apply the methodology herein to theories with more complicated symmetries.  In this story, the $\mathbb{Z}_2$ symmetry that is gauged in fermionization still plays a crucial role.  If the global symmetry is $\mathbb{Z}_2 \times H$, one still would obtain the invertible theories we discussed upon fermionization.  One could proceed analogously to \cite{Karch:2025jem}.  If the global symmetry is not a direct product of $\mathbb{Z}_2$ and some other group but instead is a nontrivial extension of $H$ by $\mathbb{Z}_2$, one would need some sort of twisted {Pin} structure.  For a recent discussion of twisted Spin structures, see \cite{Wen:2024udn}.
    \item Since gauging $\mathbb{Z}_2$ maps the Ising CFT to itself, one may view the topological interface created by gauging $\mathbb{Z}_2$ on half of spacetime as a symmetry.  It turns out that this symmetry is non-invertible \cite{Choi:2021kmx,Kaidi:2021xfk}.  It is energizing to consider similar symmetries in the theories we discuss in this paper, which we would expect to show up when a composition of the topological operations addressed exchanges the two phases in the phase diagram, analogous to how gauging $\mathbb{Z}_2$ exchanges the trivial and SSB phases.  How does the presence of time-reversal symmetry impact the physics of these symmetries and their mathematical description in terms of category theory?  Moreover, the non-invertible symmetry of the Ising CFT is related to Majorana zero modes in the Majorana CFT \cite{Thorngren:2021yso}.  Fleshing out this relationship in the presence of time-reversal symmetry seems worthwhile.
\end{itemize}
\appendix
\acknowledgments

The author thanks Andreas Karch for discussions and collaboration on a related work \cite{Karch:2025jem} and David Tong and John McGreevy for commments on the draft.  The author also thanks Jacques Distler for clarifications on the initial version.  This work is partially supported by a grant from the Simons Foundation (Grant 651678, AK).

\bibliographystyle{JHEP}
\bibliography{biblio.bib}

\providecommand{\href}[2]{#2}\begingroup\raggedright\begin{thebibliography}{10}

\bibitem{Gaiotto:2014kfa}
D.~Gaiotto, A.~Kapustin, N.~Seiberg and B.~Willett, \emph{{Generalized Global Symmetries}}, \href{https://doi.org/10.1007/JHEP02(2015)172}{\emph{JHEP} {\bfseries 02} (2015) 172} [\href{https://arxiv.org/abs/1412.5148}{{\ttfamily 1412.5148}}].

\bibitem{Cordova:2022ruw}
C.~Cordova, T.T.~Dumitrescu, K.~Intriligator and S.-H.~Shao, \emph{{Snowmass White Paper: Generalized Symmetries in Quantum Field Theory and Beyond}},  in \emph{{Snowmass 2021}}, 5, 2022 [\href{https://arxiv.org/abs/2205.09545}{{\ttfamily 2205.09545}}].

\bibitem{McGreevy:2022oyu}
J.~McGreevy, \emph{{Generalized Symmetries in Condensed Matter}}, \href{https://doi.org/10.1146/annurev-conmatphys-040721-021029}{\emph{Ann. Rev. Condensed Matter Phys.} {\bfseries 14} (2023) 57} [\href{https://arxiv.org/abs/2204.03045}{{\ttfamily 2204.03045}}].

\bibitem{Brennan:2023mmt}
T.D.~Brennan and S.~Hong, \emph{{Introduction to Generalized Global Symmetries in QFT and Particle Physics}},  \href{https://arxiv.org/abs/2306.00912}{{\ttfamily 2306.00912}}.

\bibitem{Schafer-Nameki:2023jdn}
S.~Schafer-Nameki, \emph{{ICTP lectures on (non-)invertible generalized symmetries}}, \href{https://doi.org/10.1016/j.physrep.2024.01.007}{\emph{Phys. Rept.} {\bfseries 1063} (2024) 1} [\href{https://arxiv.org/abs/2305.18296}{{\ttfamily 2305.18296}}].

\bibitem{Shao:2023gho}
S.-H.~Shao, \emph{{What's Done Cannot Be Undone: TASI Lectures on Non-Invertible Symmetry}},  \href{https://arxiv.org/abs/2308.00747}{{\ttfamily 2308.00747}}.

\bibitem{Bhardwaj:2023kri}
L.~Bhardwaj, L.E.~Bottini, L.~Fraser-Taliente, L.~Gladden, D.S.W.~Gould, A.~Platschorre et~al., \emph{{Lectures on generalized symmetries}}, \href{https://doi.org/10.1016/j.physrep.2023.11.002}{\emph{Phys. Rept.} {\bfseries 1051} (2024) 1} [\href{https://arxiv.org/abs/2307.07547}{{\ttfamily 2307.07547}}].

\bibitem{Gaiotto:2020iye}
D.~Gaiotto and J.~Kulp, \emph{{Orbifold groupoids}}, \href{https://doi.org/10.1007/JHEP02(2021)132}{\emph{JHEP} {\bfseries 02} (2021) 132} [\href{https://arxiv.org/abs/2008.05960}{{\ttfamily 2008.05960}}].

\bibitem{Karch:2019lnn}
A.~Karch, D.~Tong and C.~Turner, \emph{{A Web of 2d Dualities: ${\bf Z}_2$ Gauge Fields and Arf Invariants}}, \href{https://doi.org/10.21468/SciPostPhys.7.1.007}{\emph{SciPost Phys.} {\bfseries 7} (2019) 007} [\href{https://arxiv.org/abs/1902.05550}{{\ttfamily 1902.05550}}].

\bibitem{Karch:2025jem}
A.~Karch and R.C.~Spieler, \emph{{Critical theories connecting gapped phases with $\mathbb{Z}_2\times\mathbb{Z}_2$ symmetry from the duality web}},  \href{https://arxiv.org/abs/2502.14032}{{\ttfamily 2502.14032}}.

\bibitem{Karch:2016sxi}
A.~Karch and D.~Tong, \emph{{Particle-Vortex Duality from 3d Bosonization}}, \href{https://doi.org/10.1103/PhysRevX.6.031043}{\emph{Phys. Rev. X} {\bfseries 6} (2016) 031043} [\href{https://arxiv.org/abs/1606.01893}{{\ttfamily 1606.01893}}].

\bibitem{Seiberg:2016gmd}
N.~Seiberg, T.~Senthil, C.~Wang and E.~Witten, \emph{{A Duality Web in 2+1 Dimensions and Condensed Matter Physics}}, \href{https://doi.org/10.1016/j.aop.2016.08.007}{\emph{Annals Phys.} {\bfseries 374} (2016) 395} [\href{https://arxiv.org/abs/1606.01989}{{\ttfamily 1606.01989}}].

\bibitem{Senthil:2018cru}
T.~Senthil, D.T.~Son, C.~Wang and C.~Xu, \emph{{Duality between $(2+1)d$ Quantum Critical Points}}, \href{https://doi.org/10.1016/j.physrep.2019.09.001}{\emph{Phys. Rept.} {\bfseries 827} (2019) 1} [\href{https://arxiv.org/abs/1810.05174}{{\ttfamily 1810.05174}}].

\bibitem{Turner:2019wnh}
C.~Turner, \emph{{Dualities in 2+1 Dimensions}}, \href{https://doi.org/10.22323/1.349.0001}{\emph{PoS} {\bfseries Modave2018} (2019) 001} [\href{https://arxiv.org/abs/1905.12656}{{\ttfamily 1905.12656}}].

\bibitem{Witten:2003ya}
E.~Witten, \emph{{SL(2,Z) action on three-dimensional conformal field theories with Abelian symmetry}},  in \emph{{From Fields to Strings: Circumnavigating Theoretical Physics: A Conference in Tribute to Ian Kogan}}, pp.~1173--1200, 7, 2003 [\href{https://arxiv.org/abs/hep-th/0307041}{{\ttfamily hep-th/0307041}}].

\bibitem{TachikawaNotes}
Y.~Tachikawa, ``{Anomalies and Topological Phases}.'' \url{https://member.ipmu.jp/yuji.tachikawa/lectures/2019-top-anom/tasi2019.pdf}, 2019.

\bibitem{Kaidi:2022cpf}
J.~Kaidi, K.~Ohmori and Y.~Zheng, \emph{{Symmetry TFTs for Non-invertible Defects}}, \href{https://doi.org/10.1007/s00220-023-04859-7}{\emph{Commun. Math. Phys.} {\bfseries 404} (2023) 1021} [\href{https://arxiv.org/abs/2209.11062}{{\ttfamily 2209.11062}}].

\bibitem{Kaidi:2023maf}
J.~Kaidi, E.~Nardoni, G.~Zafrir and Y.~Zheng, \emph{{Symmetry TFTs and anomalies of non-invertible symmetries}}, \href{https://doi.org/10.1007/JHEP10(2023)053}{\emph{JHEP} {\bfseries 10} (2023) 053} [\href{https://arxiv.org/abs/2301.07112}{{\ttfamily 2301.07112}}].

\bibitem{Zhang:2023wlu}
C.~Zhang and C.~C\'ordova, \emph{{Anomalies of $(1+1)D$ categorical symmetries}},  \href{https://arxiv.org/abs/2304.01262}{{\ttfamily 2304.01262}}.

\bibitem{Antinucci:2023ezl}
A.~Antinucci, F.~Benini, C.~Copetti, G.~Galati and G.~Rizi, \emph{{Anomalies of non-invertible self-duality symmetries: fractionalization and gauging}},  \href{https://arxiv.org/abs/2308.11707}{{\ttfamily 2308.11707}}.

\bibitem{Cordova:2023bja}
C.~Cordova, P.-S.~Hsin and C.~Zhang, \emph{{Anomalies of Non-Invertible Symmetries in (3+1)d}},  \href{https://arxiv.org/abs/2308.11706}{{\ttfamily 2308.11706}}.

\bibitem{Moradi:2022lqp}
H.~Moradi, S.F.~Moosavian and A.~Tiwari, \emph{{Topological holography: Towards a unification of Landau and beyond-Landau physics}}, \href{https://doi.org/10.21468/SciPostPhysCore.6.4.066}{\emph{SciPost Phys. Core} {\bfseries 6} (2023) 066} [\href{https://arxiv.org/abs/2207.10712}{{\ttfamily 2207.10712}}].

\bibitem{Bhardwaj:2023ayw}
L.~Bhardwaj and S.~Schafer-Nameki, \emph{{Generalized Charges, Part II: Non-Invertible Symmetries and the Symmetry TFT}},  \href{https://arxiv.org/abs/2305.17159}{{\ttfamily 2305.17159}}.

\bibitem{Bhardwaj:2023idu}
L.~Bhardwaj, L.E.~Bottini, D.~Pajer and S.~Sch\"afer-Nameki, \emph{{Gapped Phases with Non-Invertible Symmetries: (1+1)d}},  \href{https://arxiv.org/abs/2310.03784}{{\ttfamily 2310.03784}}.

\bibitem{Bhardwaj:2023fca}
L.~Bhardwaj, L.E.~Bottini, D.~Pajer and S.~Schafer-Nameki, \emph{{Categorical Landau Paradigm for Gapped Phases}},  \href{https://arxiv.org/abs/2310.03786}{{\ttfamily 2310.03786}}.

\bibitem{Bhardwaj:2023bbf}
L.~Bhardwaj, L.E.~Bottini, D.~Pajer and S.~Schafer-Nameki, \emph{{The Club Sandwich: Gapless Phases and Phase Transitions with Non-Invertible Symmetries}},  \href{https://arxiv.org/abs/2312.17322}{{\ttfamily 2312.17322}}.

\bibitem{Bhardwaj:2024qrf}
L.~Bhardwaj, D.~Pajer, S.~Schafer-Nameki and A.~Warman, \emph{{Hasse Diagrams for Gapless SPT and SSB Phases with Non-Invertible Symmetries}},  \href{https://arxiv.org/abs/2403.00905}{{\ttfamily 2403.00905}}.

\bibitem{Bhardwaj:2024ydc}
L.~Bhardwaj, K.~Inamura and A.~Tiwari, \emph{{Fermionic Non-Invertible Symmetries in (1+1)d: Gapped and Gapless Phases, Transitions, and Symmetry TFTs}},  \href{https://arxiv.org/abs/2405.09754}{{\ttfamily 2405.09754}}.

\bibitem{Huang:2023pyk}
S.-J.~Huang and M.~Cheng, \emph{{Topological holography, quantum criticality, and boundary states}},  \href{https://arxiv.org/abs/2310.16878}{{\ttfamily 2310.16878}}.

\bibitem{Kong:2020cie}
L.~Kong, T.~Lan, X.-G.~Wen, Z.-H.~Zhang and H.~Zheng, \emph{{Algebraic higher symmetry and categorical symmetry -- a holographic and entanglement view of symmetry}}, \href{https://doi.org/10.1103/PhysRevResearch.2.043086}{\emph{Phys. Rev. Res.} {\bfseries 2} (2020) 043086} [\href{https://arxiv.org/abs/2005.14178}{{\ttfamily 2005.14178}}].

\bibitem{Chatterjee:2022kxb}
A.~Chatterjee and X.-G.~Wen, \emph{{Symmetry as a shadow of topological order and a derivation of topological holographic principle}}, \href{https://doi.org/10.1103/PhysRevB.107.155136}{\emph{Phys. Rev. B} {\bfseries 107} (2023) 155136} [\href{https://arxiv.org/abs/2203.03596}{{\ttfamily 2203.03596}}].

\bibitem{Chatterjee:2022tyg}
A.~Chatterjee and X.-G.~Wen, \emph{{Holographic theory for continuous phase transitions: Emergence and symmetry protection of gaplessness}}, \href{https://doi.org/10.1103/PhysRevB.108.075105}{\emph{Phys. Rev. B} {\bfseries 108} (2023) 075105} [\href{https://arxiv.org/abs/2205.06244}{{\ttfamily 2205.06244}}].

\bibitem{Kapustin:2014tfa}
A.~Kapustin, \emph{{Symmetry Protected Topological Phases, Anomalies, and Cobordisms: Beyond Group Cohomology}},  \href{https://arxiv.org/abs/1403.1467}{{\ttfamily 1403.1467}}.

\bibitem{Kapustin:2014dxa}
A.~Kapustin, R.~Thorngren, A.~Turzillo and Z.~Wang, \emph{{Fermionic Symmetry Protected Topological Phases and Cobordisms}}, \href{https://doi.org/10.1007/JHEP12(2015)052}{\emph{JHEP} {\bfseries 12} (2015) 052} [\href{https://arxiv.org/abs/1406.7329}{{\ttfamily 1406.7329}}].

\bibitem{Gaiotto:2015zta}
D.~Gaiotto and A.~Kapustin, \emph{{Spin TQFTs and fermionic phases of matter}}, \href{https://doi.org/10.1142/S0217751X16450445}{\emph{Int. J. Mod. Phys. A} {\bfseries 31} (2016) 1645044} [\href{https://arxiv.org/abs/1505.05856}{{\ttfamily 1505.05856}}].

\bibitem{Thorngren:2018bhj}
R.~Thorngren, \emph{{Anomalies and Bosonization}}, \href{https://doi.org/10.1007/s00220-020-03830-0}{\emph{Commun. Math. Phys.} {\bfseries 378} (2020) 1775} [\href{https://arxiv.org/abs/1810.04414}{{\ttfamily 1810.04414}}].

\bibitem{Bhardwaj:2016dtk}
L.~Bhardwaj, \emph{{Unoriented 3d TFTs}}, \href{https://doi.org/10.1007/JHEP05(2017)048}{\emph{JHEP} {\bfseries 05} (2017) 048} [\href{https://arxiv.org/abs/1611.02728}{{\ttfamily 1611.02728}}].

\bibitem{Debray:2018wfz}
A.~Debray and S.~Gunningham, \emph{{The Arf-Brown TQFT of Pin$^-$ Surfaces}},  \href{https://arxiv.org/abs/1803.11183}{{\ttfamily 1803.11183}}.

\bibitem{Turzillo:2018ynq}
A.~Turzillo, \emph{{Diagrammatic State Sums for 2D Pin-Minus TQFTs}}, \href{https://doi.org/10.1007/JHEP03(2020)019}{\emph{JHEP} {\bfseries 03} (2020) 019} [\href{https://arxiv.org/abs/1811.12654}{{\ttfamily 1811.12654}}].

\bibitem{Kobayashi:2019xxg}
R.~Kobayashi, \emph{{Pin TQFT and Grassmann integral}}, \href{https://doi.org/10.1007/JHEP12(2019)014}{\emph{JHEP} {\bfseries 12} (2019) 014} [\href{https://arxiv.org/abs/1905.05902}{{\ttfamily 1905.05902}}].

\bibitem{Kaidi:2019tyf}
J.~Kaidi, J.~Parra-Martinez, Y.~Tachikawa and w.a.m.a.b.A.~Debray, \emph{{Topological Superconductors on Superstring Worldsheets}}, \href{https://doi.org/10.21468/SciPostPhys.9.1.010}{\emph{SciPost Phys.} {\bfseries 9} (2020) 10} [\href{https://arxiv.org/abs/1911.11780}{{\ttfamily 1911.11780}}].

\bibitem{Bhardwaj:2020ymp}
L.~Bhardwaj, Y.~Lee and Y.~Tachikawa, \emph{{$SL(2,\mathbb{Z})$ action on QFTs with $\mathbb{Z}_2$ symmetry and the Brown-Kervaire invariants}}, \href{https://doi.org/10.1007/JHEP11(2020)141}{\emph{JHEP} {\bfseries 11} (2020) 141} [\href{https://arxiv.org/abs/2009.10099}{{\ttfamily 2009.10099}}].

\bibitem{Barkeshli:2023bta}
M.~Barkeshli, P.-S.~Hsin and R.~Kobayashi, \emph{{Higher-group symmetry of (3+1)D fermionic $\mathbb{Z}_2$ gauge theory: Logical CCZ, CS, and T gates from higher symmetry}}, \href{https://doi.org/10.21468/SciPostPhys.16.5.122}{\emph{SciPost Phys.} {\bfseries 16} (2024) 122} [\href{https://arxiv.org/abs/2311.05674}{{\ttfamily 2311.05674}}].

\bibitem{Turzillo:2023yyr}
A.~Turzillo and M.~You, \emph{{Duality and stacking of bosonic and fermionic SPT phases}}, \href{https://doi.org/10.1007/JHEP10(2024)034}{\emph{JHEP} {\bfseries 10} (2024) 034} [\href{https://arxiv.org/abs/2311.18782}{{\ttfamily 2311.18782}}].

\bibitem{Rey:2025jno}
A.~Rey, O.M.~Aksoy, D.P.~Arovas, C.~Chamon and C.~Mudry, \emph{{Incommensurate gapless ferromagnetism connecting competing symmetry-enriched deconfined quantum phase transitions}},  \href{https://arxiv.org/abs/2502.14958}{{\ttfamily 2502.14958}}.

\bibitem{Verresen:2019igf}
R.~Verresen, R.~Thorngren, N.G.~Jones and F.~Pollmann, \emph{{Gapless Topological Phases and Symmetry-Enriched Quantum Criticality}}, \href{https://doi.org/10.1103/PhysRevX.11.041059}{\emph{Phys. Rev. X} {\bfseries 11} (2021) 041059} [\href{https://arxiv.org/abs/1905.06969}{{\ttfamily 1905.06969}}].

\bibitem{Thorngren:2020wet}
R.~Thorngren, A.~Vishwanath and R.~Verresen, \emph{{Intrinsically gapless topological phases}}, \href{https://doi.org/10.1103/PhysRevB.104.075132}{\emph{Phys. Rev. B} {\bfseries 104} (2021) 075132} [\href{https://arxiv.org/abs/2008.06638}{{\ttfamily 2008.06638}}].

\bibitem{Li:2022jbf}
L.~Li, M.~Oshikawa and Y.~Zheng, \emph{{Decorated Defect Construction of Gapless-SPT States}},  \href{https://arxiv.org/abs/2204.03131}{{\ttfamily 2204.03131}}.

\bibitem{Wen:2022tkg}
R.~Wen and A.C.~Potter, \emph{{Bulk-boundary correspondence for intrinsically gapless symmetry-protected topological phases from group cohomology}}, \href{https://doi.org/10.1103/PhysRevB.107.245127}{\emph{Phys. Rev. B} {\bfseries 107} (2023) 245127} [\href{https://arxiv.org/abs/2208.09001}{{\ttfamily 2208.09001}}].

\bibitem{Li:2023ani}
L.~Li, M.~Oshikawa and Y.~Zheng, \emph{{Noninvertible duality transformation between symmetry-protected topological and spontaneous symmetry breaking phases}}, \href{https://doi.org/10.1103/PhysRevB.108.214429}{\emph{Phys. Rev. B} {\bfseries 108} (2023) 214429} [\href{https://arxiv.org/abs/2301.07899}{{\ttfamily 2301.07899}}].

\bibitem{Wen:2023otf}
R.~Wen and A.C.~Potter, \emph{{Classification of 1+1D gapless symmetry protected phases via topological holography}},  \href{https://arxiv.org/abs/2311.00050}{{\ttfamily 2311.00050}}.

\bibitem{Chen:2014xhe}
X.~Chen and A.~Vishwanath, \emph{{Towards Gauging Time-Reversal Symmetry: A Tensor Network Approach}}, \href{https://doi.org/10.1103/PhysRevX.5.041034}{\emph{Phys. Rev. X} {\bfseries 5} (2015) 041034} [\href{https://arxiv.org/abs/1401.3736}{{\ttfamily 1401.3736}}].

\bibitem{McNamara:2022lrw}
J.~McNamara and M.~Reece, \emph{{Reflections on Parity Breaking}},  \href{https://arxiv.org/abs/2212.00039}{{\ttfamily 2212.00039}}.

\bibitem{Harlow:2023hjb}
D.~Harlow and T.~Numasawa, \emph{{Gauging spacetime inversions in quantum gravity}},  \href{https://arxiv.org/abs/2311.09978}{{\ttfamily 2311.09978}}.

\bibitem{Fidkowski:2010jmn}
L.~Fidkowski and A.~Kitaev, \emph{{Topological phases of fermions in one dimension}}, \href{https://doi.org/10.1103/PhysRevB.83.075103}{\emph{Phys. Rev. B} {\bfseries 83} (2011) 075103} [\href{https://arxiv.org/abs/1008.4138}{{\ttfamily 1008.4138}}].

\bibitem{BoyleSmith:2024qgx}
P.~Boyle~Smith and Y.~Zheng, \emph{{Backfiring Bosonisation}},  \href{https://arxiv.org/abs/2403.03953}{{\ttfamily 2403.03953}}.

\bibitem{Witten:2015aba}
E.~Witten, \emph{{Fermion Path Integrals And Topological Phases}}, \href{https://doi.org/10.1103/RevModPhys.88.035001}{\emph{Rev. Mod. Phys.} {\bfseries 88} (2016) 035001} [\href{https://arxiv.org/abs/1508.04715}{{\ttfamily 1508.04715}}].

\bibitem{Wen:2024udn}
R.~Wen, W.~Ye and A.C.~Potter, \emph{{Topological holography for fermions}},  \href{https://arxiv.org/abs/2404.19004}{{\ttfamily 2404.19004}}.

\bibitem{Choi:2021kmx}
Y.~Choi, C.~Cordova, P.-S.~Hsin, H.T.~Lam and S.-H.~Shao, \emph{{Noninvertible duality defects in 3+1 dimensions}}, \href{https://doi.org/10.1103/PhysRevD.105.125016}{\emph{Phys. Rev. D} {\bfseries 105} (2022) 125016} [\href{https://arxiv.org/abs/2111.01139}{{\ttfamily 2111.01139}}].

\bibitem{Kaidi:2021xfk}
J.~Kaidi, K.~Ohmori and Y.~Zheng, \emph{{Kramers-Wannier-like Duality Defects in (3+1)D Gauge Theories}}, \href{https://doi.org/10.1103/PhysRevLett.128.111601}{\emph{Phys. Rev. Lett.} {\bfseries 128} (2022) 111601} [\href{https://arxiv.org/abs/2111.01141}{{\ttfamily 2111.01141}}].

\bibitem{Thorngren:2021yso}
R.~Thorngren and Y.~Wang, \emph{{Fusion category symmetry. Part II. Categoriosities at c = 1 and beyond}}, \href{https://doi.org/10.1007/JHEP07(2024)051}{\emph{JHEP} {\bfseries 07} (2024) 051} [\href{https://arxiv.org/abs/2106.12577}{{\ttfamily 2106.12577}}].

\end{thebibliography}\endgroup


\end{document}